\documentclass[prl,twocolumn,showpacs,superscriptaddress,nofootinbib]{revtex4-1}
\usepackage{graphicx}
\usepackage{bm}
\usepackage{amssymb}
\usepackage{color}
\usepackage{amsmath}
\usepackage{enumerate}
\usepackage{amstext}
\usepackage{latexsym}
\usepackage[usenames,dvipsnames]{xcolor}
\usepackage[colorlinks=true,citecolor=Cerulean,linkcolor=RubineRed,urlcolor=Cerulean]{hyperref}
\usepackage{multibib}
\usepackage{mathptmx} 
\newcommand{\pac}[1]{ \left\{ #1 \right\} }
\newcommand{\pap}[1]{\left( #1 \right)}

\newcommand{\bra}[1]{\left\langle #1 \right\vert}

\newcommand{\ket}[1]{\left\vert #1 \right\rangle}

\newcommand{\beq}{\begin{equation}}
\newcommand{\eeq}{\end{equation}}
\newcommand{\beqa}{\begin{eqnarray}}
\newcommand{\eeqa}{\end{eqnarray}}

\begin{document}

\title{Universal Dynamics of  Inhomogeneous Quantum Phase Transitions: Suppressing Defect Formation}
\author{F. J. G\'omez-Ruiz}
\affiliation{Departamento de F{\'i}sica, Universidad de los Andes, A.A. 4976, Bogot\'a D. C., Colombia}
\affiliation{Department of Physics, University of Massachusetts, Boston, MA 02125, USA}
\author{A. del Campo}
\affiliation{Donostia International Physics Center,  E-20018 San Sebasti\'an, Spain}
\affiliation{IKERBASQUE, Basque Foundation for Science, E-48013 Bilbao, Spain}
\affiliation{Department of Physics, University of Massachusetts, Boston, MA 02125, USA}
\affiliation{Theoretical Division, Los Alamos National Laboratory, MS-B213, Los Alamos, NM 87545, USA}

\begin{abstract}
In the nonadiabatic dynamics across a quantum phase transition,  the Kibble-Zurek mechanism predicts that the formation of topological defects is suppressed as a universal power law with the quench time. In inhomogeneous systems, the critical point is reached locally and causality reduces the effective system size for defect formation to regions where the velocity of the critical front is slower than the sound velocity, favoring adiabatic dynamics. The reduced density of excitations  exhibits a much steeper dependence on the quench rate and is also described by a universal power-law, that we demonstrated in a quantum Ising chain.
\end{abstract}
\maketitle
The development of new methods to induce or mimic adiabatic dynamics is essential to the progress of quantum technologies. In many-body systems, the need to develop new methods to approach adiabatic dynamics is underlined for their potential application to quantum simulation and adiabatic quantum computation~\cite{QS82,CZ12}. 

The Kibble-Zurek mechanism (KZM) is a paradigmatic theory to describe the dynamics across both classical continuous phase transitions and quantum phase transitions~\cite{Kibble76a,Kibble76b, Zurek96a,*Zurek96b,*Zurek96c, Dziarmaga10,Polkovnikov11}. The system of interest  is assumed to be driven by a quench of an external control parameter $h(t)=h_c(1-t/\tau_Q)$ in a finite time $\tau_Q$ across the critical value $h_c$. The mechanism exploits the divergence of the relaxation time $\tau(\epsilon)=\tau_0/|\epsilon|^{z\nu}$ (critical slowing down) as a function of the dimensionless distance  to the critical point  $\epsilon=(h_c-h)/h_c=t/\tau_Q$, to estimate the time scale, known as the freeze-out time $\hat{t}$, in which the dynamics ceases to be adiabatic.  The dynamics is therefore controlled by the quench time $\tau_Q$ and by  $z$ and $\nu$,  which are referred to as   the dynamic and correlation-length critical exponent, respectively. The central prediction of the KZM is the estimate of the size of the domains in the broken symmetry phase using the equilibrium value of the correlation length $\xi(\epsilon)=\xi_0/|\epsilon|^{\nu}$, at the value $\epsilon(\hat{t})=\hat{\epsilon}$. As a result, the average domain size exhibits a universal power-law scaling dictated by $\xi(\hat{t})=\xi_0(\tau_Q/\tau_0)^{\nu/(1+z\nu)}$.  At the boundary between domains, topological defects form. In one dimension, the density of defects is set by $d=\xi(\hat{t})^{-1}\sim\tau_Q^{-\beta_{\rm KZM}}$ with $\beta_{\rm KZM}=\nu/(1+z\nu)$. The KZM constitutes a negative result for the purpose of suppressing defect formation, given that in an arbitrarily large system, defects will be formed no matter how slowly the phase transition is crossed. This has motivated  a variety of approaches to circumvent the KZM scaling law and favor adiabatic dynamics, including  nonlinear protocols \cite{Diptiman08,Barankov08},   optimal control \cite{Doria11,Rahmani11,DeChiara13}, shortcuts to adiabaticity \cite{delcampo12,Saberi14,Campbell15}, and the simultaneous tuning of multiple parameters of the system \cite{SauSengupta14},  to name some relevant examples \cite{DS15}.
 
Test-beds for the experimental demonstration of universal dynamics at criticality are often inhomogeneous, and it is this feature which paves the way to defect suppression. Under a finite-rate quench of an external control parameter, the system does not reach the critical point everywhere at once. Rather,  a choice of the broken symmetry  made locally at the critical front can influence the subsequent symmetry breaking across the system, diminishing the overall number of defects. In this scenario, the paradigmatic KZM fails, and should be extended to account for the inhomogeneous character of the system~\cite{ZD08, Zurek09, DM10, DM10b, ions1, DRP11}. An Inhomogeneous Kibble-Zurek mechanism (IKZM) has been formulated in classical phase transitions~\cite{Zurek09, ions1, DRP11} following the early insight by Kibble and Volovik~\cite{KV97}. The current understanding is summarized in~\cite{DKZ13,DZ14}. Its key predictions are a suppression of the net number  of excitations with respect to the homogeneous scenario, and an enhanced power-law scaling of the residual density of excitations as a function of the quench rate.

In classical systems, numerical evidences in favor of the IKZM have been reported ~\cite{ions1}. Three experimental groups have observed an enhanced dependence of the density of kinks with the quench rate  across a structural continuous phase transition in trapped Coulomb crystals ~\cite{EH13,Ulm13,Pyka13}.  However, a related experiment testing soliton formation during Bose-Einstein condensation of a trapped atomic cloud  under forced evaporative cooling was consistent with the standard KZM in a homogeneous setting~\cite{Lamporesi13}. In addition,  a verification of the power-law in both numerical studies and experiments has been limited by the range of testable quench rates and defect losses. Defect suppression induced by causality has also been shown to play a role in  inhomogeneous quantum systems, that have so far been explored by numerics and adiabatic perturbation theory~\cite{CK10,DM10,DM10b,Rams16,Nishimori18,Mohseni18}. 

In this work, we establish the universal character of the critical dynamics across an inhomogeneous quantum phase transition and the validity of the IKZM in the quantum domain. We show that the dependence of the density of excitations with the quench rate  is universal and exhibits a crossover between the standard KZM at fast quench rates, and a steeper power-law dependence for slower ramps, that favors defect suppression. 

\begin{figure}[t]
\begin{center}
\includegraphics[scale=0.75]{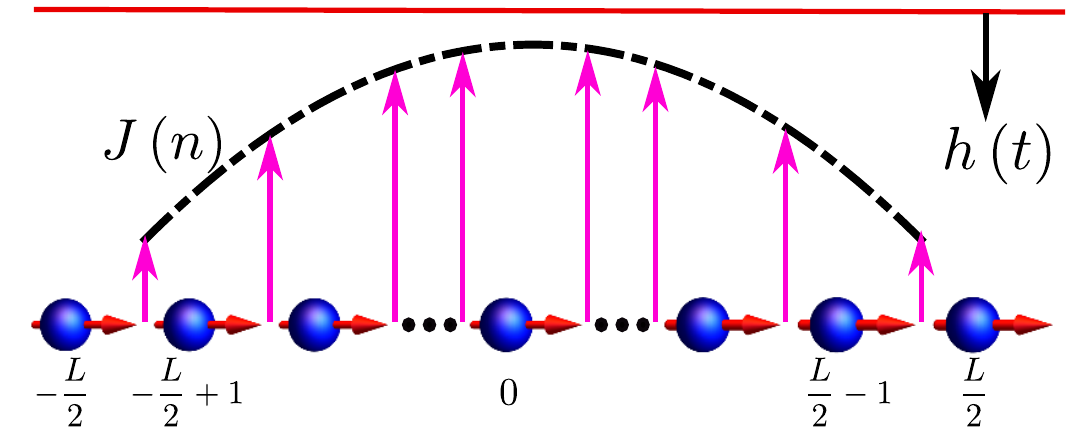}
\caption{{\bf Inhomogeneous quantum phase transition.} Schematic illustration of a one-dimensional transverse-field quantum Ising chain with a symmetric spatial modulation of the tunneling amplitude  $J\pap{n}$ (vertical arrow). As the homogeneous magnetic field $h(t)$ (red line) is decreased, the critical point is first crossed locally at the center of the chain. Subsequently, the critical front spreads sideways at a speed $v_F(n)$ that can be controlled by the quench rate. 
}\label{fig_1}
\end{center}
\end{figure}

{\it Dynamics of an inhomogeneous quantum phase transition.---}The one-dimensional inhomogeneous quantum Ising model in a transverse magnetic field $h$ describes a chain of $L$ spins with the Hamiltonian
\beqa
\mathcal{H}_0=-\sum_{n=1}^{L-1} J(n)\sigma_{n}^{z}\sigma_{n+1}^{z}-\sum_{n=1}^{L}h\pap{t} \sigma_{n}^{x}.
\label{H_Ising}
\eeqa
The setup~\eqref{H_Ising} is schematically represented in Figure \ref{fig_1}. 

Its homogeneous version ($J(n)=J$) is a paradigmatic model to study quantum phase transitions~\cite{Sachdev},  and its quantum simulation in the laboratory is at reach in a variety of quantum platforms including superconducting circuits \cite{Barends2016a}, Rydberg atoms \cite{Labuhn2016} and trapped ions \cite{Monroe17}. 
The homogeneous transverse-field Ising model (H-TFIM) exhibits a quantum phase transition at $h_c = \pm J$ between a paramagnetic phase ($|h|>J$) and a ferromagnetic phase ($|h|<J$).  Therefore, it is convenient to introduce the reduced parameter $\varepsilon=(J-h)/J$. The gap between the ground and excite state closes as $\Delta= 2|h-J|$, so the relaxation time $\tau=\hbar/\Delta=\tau_0/|\varepsilon|$ diverges as the system approaches the critical point (critical slowing down). The equilibrium healing length reads $\xi=2J/\Delta=1/|\varepsilon|$ in units of the lattice spacing. 

The critical dynamics of a  H-TFIM is well-described by the standard KZM \cite{Dziarmaga10,Polkovnikov11}. The nonadiabatic dynamics results in the creation of topological defects. In the classical case, the latter are formed at the boundary between adjacent domains in the broken symmetry phase and are known as ($\mathbb{Z}_2$) kinks. In the quantum domain, excitations involve coherent quantum superpositions and are generally  delocalized \cite{Dziarmaga12}. This is particularly the case in translationally invariant systems \cite{Dziarmaga05}.
The quantum KZM sets the average distance between kinks by the equilibrium value of the correlation length at the instant when the dynamics ceases to be adiabatic~\cite{Polkovnikov05,Damski05,Dziarmaga05,ZDZ05}. This time scale known as the freeze-out time can be estimated by equating the relaxation time to the time elapsed after the critical point, $\tau(\hat{t})=|\varepsilon/\dot{\varepsilon}|$, 
whence it follows that $\hat{t}=\sqrt{\tau_0\tau_Q}$. The density of topological defects $d\sim\xi(\hat{t})^{-1}$ scales then as 
\beqa\label{KZM}
d_{\text{KZM}}=\frac{1}{\sqrt{2J\tau_Q/\hbar}} .
\eeqa

An exact calculation shows that $d=\frac{1}{2\pi}d_{\text{KZM}}$~\cite{Dziarmaga05}. We wish to investigate how this paradigmatic scenario is modified 
in inhomogeneous quantum phase transitions, extending in doing so the IKZM to the quantum domain. We consider a smooth spatial modulation of the  tunneling amplitude $J(n)$ with maximum at $n=0$ and refer to  (\ref{H_Ising}) by I-TFIM in this case.
Using a Taylor series expansion, $J(n)$ can be locally approximated by  a quadratic function  of the form
\beqa\label{mod_J}
J(n)=J(0)(1-\alpha n^2)+\mathcal{O}(n^3).
\eeqa
The choice of $\alpha$ is such that the  interaction  coupling at the edges of the chain recovers the value in the homogeneous Ising model $J(n)=J\sim1$, that we use as a reference.  
This parabolic spatial modulation  often arises  in trapped systems using the local density approximation \cite{DKZ13} and it is accessible in quantum simulators \cite{Barends2016a,Labuhn2016,Monroe17}.
Further, we let  the quench of the  magnetic field to be homogeneous and with constant rate $1/\tau_Q$, 
\beqa
h(t)=J(0)\left(1-\frac{t}{\tau_Q}\right),
\eeqa 
during the time interval  $t\in[-\tau_Q,\tau_Q]$. 
Alternatively one could consider the driving of a homogeneous system with a spatially-dependent magnetic field.
We introduce the dimensionless control parameter  $\varepsilon(n,t)=\frac{h(t)-J(n)}{J(n)}$ that provides a notion of local  distance to the critical point. It takes values $\varepsilon(x,t)>0$ in the high symmetry (paramagnetic)  phase, reaches $\varepsilon(x,t)=0$ at the critical point, and the broken-symmetry phase for $\varepsilon(x,t)<0$ (ferromagnetic phase). In what follows, we consider the case in which the  system is initially prepared deep in the ground state of the paramagnetic phase such that $\varepsilon(n,t)>0$ everywhere in the chain.

As a result of the spatial modulation of $J(n)$, we introduce an effective quench rate with a spatial dependence,  $\tau_Q(n)=\tau_Q\frac{J(n)}{J(0)}=\tau_Q(1-\alpha n^2)$. From the condition  $\varepsilon(x_F,t_F)=0$, the time at which criticality is reached at the site $n_F=n$ is  $t_F=\tau_Q\alpha n^2$ and thus $\varepsilon(n,t)=[t-t_F(n)]/\tau_Q(n)$.  This expression determines the trajectory of the critical front  and yields the following estimate for the local front velocity
\beqa
\label{frontv}
v_F(n)=\frac{1}{2\alpha |n|\tau_Q(0)},
\eeqa
which diverges as the lattice index approaches the extrema of $J(n)$, i.e., at the center of the chain. The critical dynamics is expected to be nonadiabatic whenever $v_F(n)$ surpasses the speed of sound $s=2J(n)/\hbar$. Assuming that $J(n)\approx J(0)$ in the region where $v_F(n)>s$,   the effective size of the system  where excitations are expected to be created  is set by
\beqa
|\hat{n}|<\frac{\hbar}{4\alpha J(0)\tau_Q},
\eeqa
and is thus proportional to the quench rate.
The effective size of the system for kink formation is simply $2|\hat{n}|$ and one can thus expect a suppression of the total number of defects by a factor $2|\hat{n}|/L$ with respect to the homogeneous scenario ($J(n)=J(0)$).  The net density of defects is estimated to be given by
\beqa
d\sim\frac{2|\hat{n}|}{L\xi(\hat{t})}.
\eeqa
Assuming $\alpha|\hat{n}|^2\ll1$, one can use the estimate of the homogeneous KZM for the average distance between kinks $\xi(\hat{t})$. As a result, the estimate of the IKZM for the net number of defects in the I-TFIM reads
\beqa\label{inhomo}
d_{\text{IKZM}}=\frac{1}{\alpha L}\bigg[\frac{\hbar}{2J(0)\tau_Q}\bigg]^{\frac{3}{2}}.
\label{dikzm}
\eeqa
\begin{figure}[t!]
\begin{center}
\includegraphics[scale=0.65]{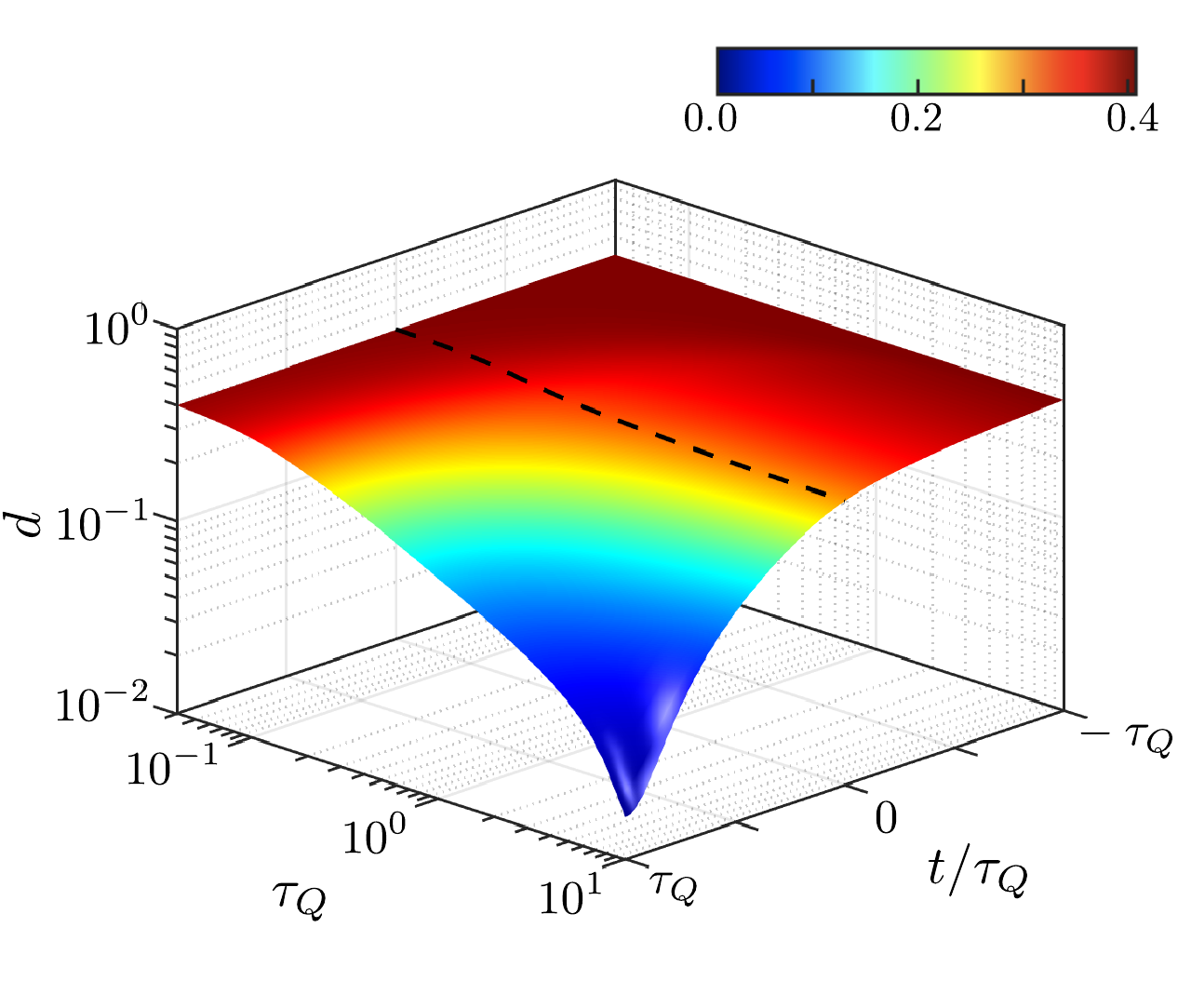}
\caption{{\bf Generation of topological defects under inhomogeneous driving.} Density of  kinks $d$ as a function of  the quench rate $\tau_Q$ and the scaled time of evolution $t/\tau_Q$. The dashed black line represent the time $t_c \pap{0}$ at which criticality is reached at the center of the linear  chain ($L=50$). The initial magnetic field is $h(-\tau_Q)=10J$ so that the initial state is deep in the paramagnetic phase, with  $J(0)=5J$ and  $J(\pm L/2)=J$ ($q=2$). }\label{fig_2}
\end{center}
\end{figure}

We note  the three-fold enhancement of  the power-law exponent, an easily testable prediction that we demonstrate numerically in what follows. The condition $\alpha\hat{n}^2\ll 1$ is not sufficient to test the IKZM scaling law  Eq. (\ref{dikzm}). When  $2\hat{n}/\hat{\xi}\sim1$ the applicability of the KZM can be called into question. Numerical simulations \cite{ions1} and several experiments \cite{EH13,Ulm13,Pyka13} have reported a steepening of the scaling at the onset of adiabatic dynamics in the course of classical phase transitions. To avoid running into this regime, we  demand $2\hat{n}/\hat{\xi}>1$.

Further,  $2\hat{n}$ should be large enough so that the power law scaling can be observed, without saturation at fast quench rates. Hence, we are led to consider slow quenches in large system sizes with small $\alpha$, and $\hbar\alpha/(4J(0))\ll\tau_Q<\hbar/(2J(0)\alpha^{2/3})$.

This treatment of the inhomogeneous quantum phase transition admits a straightforward extension to general modulations of the form $J(n)=J(0)(1-\alpha_q |n|^q)$ as shown in~\cite{*[{See the Supplemental Material at }] [{for details of the calculations and derivations.}] SM_Gomez}. The power-law dependence of the density of topological defects is then
\beqa
\label{qdikzm}
d_{\rm IKZM}
=\frac{2}{L}\left(\frac{1}{\alpha_q q}\right)^{\frac{1}{q-1}}\left(\frac{\hbar}{2J(0)\tau_Q}\right)^{\frac{q+1}{2q-2}},
\eeqa
that generalizes Eq. (\ref{dikzm}) to values of $q$ different from  $q=2$.

In order to provide quantitative evidence of the IKZM, we perform numerical simulations based on tensor-network algorithms~\cite{tnt}. Specifically, at $t=-\tau_Q$ we calculate the ground state for~\eqref{H_Ising} by means of the density matrix renormalization group (DMRG) algorithm~\cite{white1992prl, *white1993prb},  using a matrix product state description~\cite{schollwock2011ann}. This is taken as the initial state at the beginning of the quench. Subsequently we simulate the time evolution described by the time-dependent Schr\"odinger equation generated by the Hamiltonian~\eqref{H_Ising} using the time evolving block decimation (TEBD) algorithm~\cite{vidal2004prl}; \textcolor{blue}{see \cite{SM_Gomez}.} We calculate the expectation value of the operator associated with the density of kinks  
\begin{equation}\label{kinks}
d\pap{t}\equiv \frac{1}{2L}\sum_{n=1}^{L-1}\bra{\psi(t)}(1-\sigma_{n}^{z}\sigma_{n+1}^{z})\ket{\psi(t)}  
\end{equation}
where $\ket{\psi(t)}$ is the time-dependent state.  Using the direct evaluation of Eq.~\eqref{kinks}, Figure~\ref{fig_2} shows the dynamics of the density of kinks as a function of the time of evolution $t/\tau_Q$ during the crossing of the phase transition for different quench times $\tau_Q$. In particular, we consider a linear chain of $L=50$ spins with open boundary conditions, described by matrix product states with bond dimension up to $\chi=1500$. The dashed  line signals the time at which the phase transition is reached in the center of  the  chain. 
\begin{figure}[t!]
\begin{center}
\includegraphics[scale=0.7]{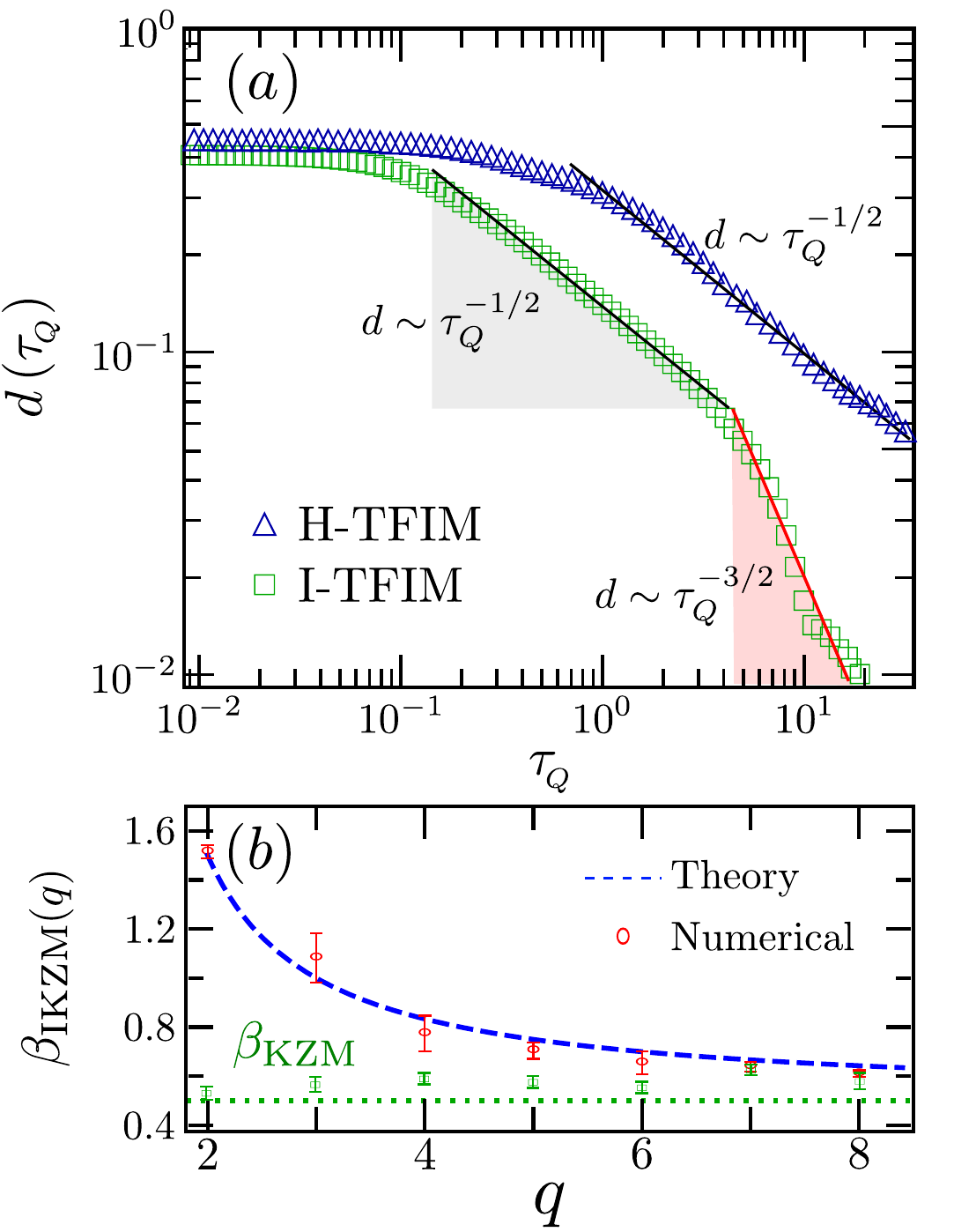}
\caption{{\bf Density of topological defects $d$ as a function of the quench time $\tau_Q$.} 
(a) In a homogeneous  transition (H-TFIM, denoted by $\triangle$) the dependence of $d$ on $\tau_{Q}$ is described by the original KZM, up to a saturation of $d$ at fast quench rates and the onset of adiabatic dynamics for slow quenches. When the critical point is crossed locally (I-TFIM, denoted by  $\Box$), the dependence is no longer described by a single-power law and exhibits a crossover between two universal regimes, described by Eqs.~\eqref{KZM} and  Eq.~\eqref{dikzm}. The symbols correspond to the numerical results for $L=50$, and the solid lines are the linear fits ($q=2$). (b) Comparison between the theoretical and numerical power-law exponents for different values of $q$ in both homogenous and inhomogeneous  scaling regimes. Results corresponding to the homogenous scaling ($q\to\infty$) are shown in green.} \label{fig_3}
\end{center}
\end{figure}

{\it Universal scaling of the density of kinks.---} The correlation-length and dynamic critical exponents of the Ising model are known to be $\nu=1$ and $z=1$, respectively. For arbitrary values $\nu$ and $z$, 
one can extend the IKZM to the quantum case  estimating the sound velocity by the ratio of the local frozen-out correlation length $\hat{\xi}(n)$ and relaxation time scale $\hat{\tau}(n)=\tau[\hat{\varepsilon}]=\hat{t}(n)$,
this is,  by 
\beqa 
\label{hats}
\hat s =  \frac {\hat \xi} {\hat \tau} = \frac {\xi_0} {\tau_0} \bigg[\frac {\tau_0} {\tau_Q(n)} \bigg]^{\frac {\nu(z-1)} {1+\nu z}}.
\eeqa  
Using $\tau_{Q}(n)\approx\tau_Q(0)$ in the resulting expression for the effective size of the system one finds in one dimension $d_{\text{IKZM}}\sim\tau_Q^{-\beta_{\rm IKZM}} $, where $\beta_{\rm IKZM}={\frac{1+2z}{1+z\nu}}$, which reduces for the TFIM ($z=\nu=1$) to $\beta_{\rm IKZM}=3/2$. For a general value of $q$, the exponent reads $\beta_{\rm IKZM}={\frac{1+q\nu}{(1+z\nu)(q-1)}}$.  Further, one can generally write 
\beqa
d_{\rm IKZM}\sim \hat{X}d_{\rm KZM}, 
\eeqa
this is, the IKZM scaling follows from the paradigmatic KZM result for homogeneous systems, taking into account the effective fraction of the system $\hat{X}=2\hat{n}/L$ that depends on the quench rate as $\hat{X}\sim 1/\tau_Q$. In Figure ~\ref{fig_3}(a), we depict the power-law dependence for the density of excitations upon completion of the phase transition ($t=\tau_Q$) as a function of $\tau_Q$ for both H-TIFM and I-TFIM cases. The homogeneous case \cite{Dziarmaga05} is governed by the universal  KZM scaling $d\sim \tau_{Q}^{-0.51\pm0.03}$, with regression coefficient $0.9994$, and is plotted as a reference. The numerical data is well-described by the  KZM prediction $d\sim 1/\sqrt{\tau_Q}$. The density of defects   also exhibits  a saturation at fast quenches as well as deviations at the onset of adiabatic dynamics in the limit of slow driving. These features  in the extreme case of very fast and slow quenches are  also shared by the inhomogeneous crossing of the phase transition.  In addition, numerical simulations in Fig.~\ref{fig_3}(a) establish that for intermediate quench rates the density of defects exhibits a crossover between two different universal regimes, dictated by the
 KZM (fast rate quench) and IKZM (slow rate quench) scalings derived in Eqs.~\eqref{KZM} and~\eqref{inhomo} respectively. We report in the I-TFIM a critical exponent for KZM of $d\sim \tau_{Q}^{-0.52\pm0.03}$ with regression coefficient $0.9991$.  Therefore, at fast quench rates the inhomogeneous critical dynamics is  described by the KZM prediction $d\sim 1/\sqrt{\tau_Q}$. Further, the scaling for slower quench rates is described by $d\sim \tau_{Q}^{-1.51\pm0.03}$ with regression coefficient $0.9992$, in  agreement with the theoretical prediction $d\sim \tau_Q^{-3/2}$. This IKZM scaling holds for quench rates between the onset fo adiabatic dynamics and he characteristic quench rate $\tau_Q^*$ at which the crossover occurs. The latter  can be estimated by setting the effective system size  equal to the physical system size, i.e., $2\hat{n}=L$. This yields $\tau_Q^*=\hbar/(2\alpha J(0)L)$ and we have verified numerically the inverse linear dependence of the  value of $\tau_Q^*$  observed numerically with the system size  $L$ and  and  $\alpha$. Figure 3(b) shows the agreement between the theoretical an analytical power-law exponents in the inhomogeneous scaling regime when different values of  $q$  are considered (other than $q=2$), governing the modulation of $J(n)=J(0)(1-\alpha_q |n|^q)$. See~\cite{SM_Gomez} for the corresponding numerical results  on the explicit dependence of the density of defects as a function of $\tau_Q$ analogous to Fig. \ref{fig_3}(a).  The agreement indicates the validity of the generalized power-law prediction in Eq. (\ref{qdikzm}). As the value of $q$ increases the nature of the transition becomes increasingly homogenous. For large values of $q$ the exponent approaches the homogenous value $\beta_{\rm KZM}=1/2$. 

{\it Conclussions.---}
In summary, we have explored the effect of local driving in the universal dynamics across a quantum phase transition using the paradigmatic quantum Ising chain as a testbed. A local crossing of the critical point can result from inhomogeneities  in  the system or the spatial modulation in the external fields that drive the transition.  As the critical point is reached locally, there is an interplay between the speed of sound and the velocity of propagation of the critical front.  The effective system size in which topological defects can form acquires then a dependence on the quench rate.  For fast quenches, the residual density of defects is well described by a power law in agreement with the original Kibble-Zurek mechanism. As the quench rate decreases there exists a crossover  to a novel power-law scaling behavior of the density of defects, that is characterized by a larger exponent, higher than that predicted by the Kibble-Zurek mechanism.  Local driving  thus leads to a much more pronounced  suppression of the density of defects, that constitute a testable prediction amenable to a variety of platforms for quantum simulation including cold atoms in optical lattices, as well as deviations at the onset of adiabatic dynamics in the limit of slow drivingtrapped ions and superconducting qubits. Our results should prove useful in a variety of contexts including the preparation of  phases of matter in quantum simulators and the engineering of inhomogeneous schedules in quantum annealing.

{\it Acknowledgment.-} It is a pleasure to acknowledge discussions with  L.-P. Garc\'ia-Pintos, J. J. Mendoza-Arenas, A. Polkovnikov, M. Rams, and D. Tielas.
F. J. G.-R. acknowledges  financial support from Facultad de Ciencias at Universidad de Los Andes (2018-II) and  thanks the University of Massachusetts Boston for hospitality during the completion of this work.
Funding support from the John Templeton Foundation and UMass Boston (project P20150000029279) is further acknowledged.
\bibliography{QIKZM_Bib}	

\begin{thebibliography}{51}
\bibitem [{\citenamefont {url will be inserted~by publisher}()}]{SM_Gomez}%
  \BibitemOpen
  \bibfield  {author} {\bibinfo {author} {\bibnamefont {url will be inserted~by
  publisher}},\ }\href@noop {} {\ }\BibitemShut {NoStop}%
\end{thebibliography}
\newpage
\pagebreak
\clearpage
\widetext
\begin{center}
\textbf{\large ---Supplemental Material---\\
 Universal Dynamics of  Inhomogeneous Quantum Phase Transitions: Suppressing Defect Formation}\\
\vspace{0.5cm}
F. J. G\'omez-Ruiz$^{1,2}$ and A. del Campo$^{3,4,2,5}$\\
\vspace{0.2cm}
$^1${\it Departamento de Física, Universidad de los Andes, A.A. 4976, Bogotá D. C., Colombia.}\\
$^2${\it Department of Physics, University of Massachusetts, Boston, MA 02125, USA}\\
$^3${\it Donostia International Physics Center,  E-20018 San Sebasti\'an, Spain}\\
$^4${\it IKERBASQUE, Basque Foundation for Science, E-48013 Bilbao, Spain}\\
$^5${\it Theoretical Division, Los Alamos National Laboratory, MS-B213, Los Alamos, NM 87545, USA}
\end{center}
\begin{center}
\vspace{1.3cm}
{\bf Contents}
\end{center}
\begin{enumerate}
\itemsep-0.5em 
\item[\textcolor{RubineRed}{\bf I.}] \textcolor{RubineRed}{\bf Density of defects with arbitrary spatial dependence of the critical front: quantum Ising model}\hfill\textcolor{RubineRed}{1}
\begin{enumerate}
\itemsep-0.5em 
       \item[\textcolor{RubineRed}{A.}] \textcolor{RubineRed}{Density of defects with arbitrary spatial dependence of the critical front and critical exponents $\nu$ and $z$}\hfill\textcolor{RubineRed}{2} 
       \item[\textcolor{RubineRed}{B.}] \textcolor{RubineRed}{Deviations from power-law behavior: quantum Ising model}\hfill\textcolor{RubineRed}{3}
\end{enumerate}  
\item[\textcolor{RubineRed}{\bf II.}] \textcolor{RubineRed}{\bf Numerical Approach}
\hfill\textcolor{RubineRed}{3}     

\begin{enumerate}
\itemsep-0.5em 
\item[\textcolor{RubineRed}{A.}] \textcolor{RubineRed}{Numerical Methods: DMRG and TEBD}\hfill\textcolor{RubineRed}{3}
\item[\textcolor{RubineRed}{B.}] \textcolor{RubineRed}{Numerical results with arbitrary spatial dependence of the critical front: quantum Ising model}\hfill\textcolor{RubineRed}{4}
\end{enumerate}
\item[] \textcolor{RubineRed}{{\bf References}}\hfill\textcolor{RubineRed}{6} 
\end{enumerate}

\setcounter{equation}{0}
\setcounter{figure}{0}
\setcounter{table}{0}
\setcounter{section}{0}
\setcounter{page}{1}
\makeatletter
\renewcommand{\theequation}{S\arabic{equation}}
\renewcommand{\thefigure}{S\arabic{figure}}
\renewcommand{\bibnumfmt}[1]{[S#1]}
\renewcommand{\citenumfont}[1]{S#1}

\section{I. Density of defects with arbitrary spatial dependence of the critical front: quantum Ising model}
Consider a  modulation of the tunneling matrix element $J(n)$ that is spatially symmetric and has a maximum at the center of chain $(n=0)$. In the main text, we have discussed the critical dynamics whenever $J(n)$ can be approximated by its truncated Taylor series expansion, 
\begin{equation}
J(n)=J(0)\left(1-\alpha n^2\right),
\end{equation}
where we disregard higher order corrections $\mathcal{O}(n^3)$ and the coefficient 
\beqa
\alpha=-\frac{J''(0)}{2J(0)}.
\eeqa
In what follows we consider the case in which $J''(0)=0$ and 
\beqa
\label{Smod_J}
J(n)=J(0)\left(1-\alpha_q |n|^q\right).
\eeqa
To keep the same end values of $J(n)$ at $n=\pm L/2$ for different values of $q$, the constant $\alpha_q$ includes a dependence on $q$. For two different values of   $q=q_1,q_2$, the corresponding constants are related by $\alpha_{q_1}=\alpha_{q_2}|L/2|^{q_2-q_1}$.

\begin{figure}[t]
\begin{center}
\includegraphics[scale=0.87]{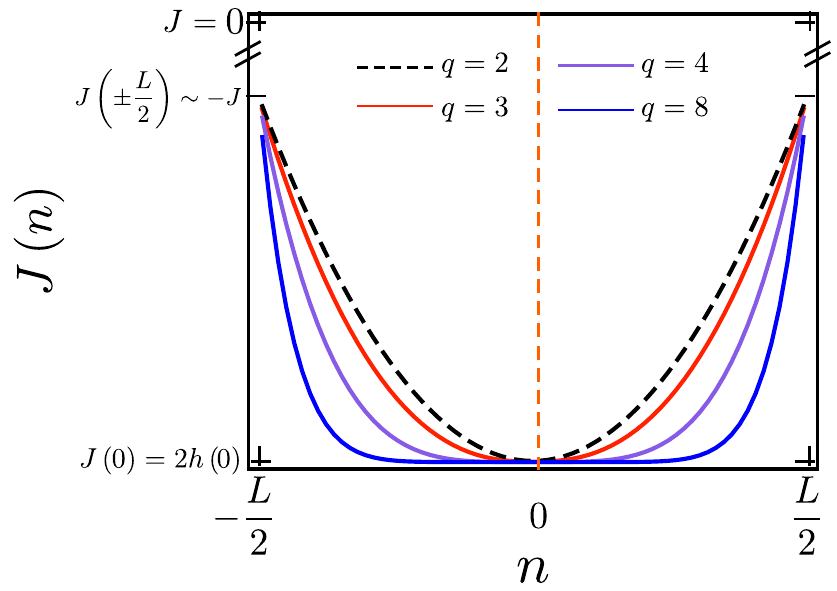}
\caption{{\bf Inhomogeneous tunneling matrix element.} Schematic illustration of a symmetric spatial modulation of the tunneling amplitude  $J\pap{n}$ given by Eq.~\eqref{Smod_J}  for differences  values of $q$. }\label{Sfig_1}
\end{center}
\end{figure}

In the figure~\ref{Sfig_1}, we show the symmetric spatial modulation considered in Eq.~\eqref{Smod_J}. The formalism in the main text can be extended to account for it by noting that the relevant front velocity is given by
\beqa
v_F=\frac{1}{\alpha_q q\tau_Q|n|^{q-1}}.
\eeqa
The divergence of the front velocity in the neighborhood  of $n=0$ is consistent with the fact that the modulation of $J(n)$  for $q>1$ is flattened and the quench is locally homogeneous ($v_F\rightarrow\infty$)  in this region.

Topological defects may only form in regions where the front velocity $v_F$ surpasses the speed of sound $s(n)=2J(n)/\hbar$. Assuming that this condition is only satisfied around $n=0$, the speed of sound can be approximated by the constant $s(n)\approx s(0)$ and the  the half-size of such region can be estimated as
\beqa
|\hat{n}|<\left(\frac{\hbar}{2\alpha_q q\tau_QJ(0)}\right)^{\frac{1}{q-1}},
\eeqa 
whenever $q>1$. The density of topological defects is then given by
\beqa
\begin{split}
\label{qdikzm}
d_{\rm IKZM}&=\frac{2|\hat{n}|}{L\hat{\xi}}\\
&=\frac{2}{L}\left(\frac{1}{\alpha_q q}\right)^{\frac{1}{q-1}}\left(\frac{\hbar}{2J(0)\tau_Q}\right)^{\frac{q+1}{2q-2}}.
\end{split}
\eeqa
The density of defects (\ref{qdikzm}) thus scales as a  power-law  $d_{\rm IKZM}\sim\tau_{Q}^{-\beta_{\rm IKZM}(q)}$ in which the power-law exponent  $\beta_{\rm IKZM}(q)=\frac{q+1}{2q-2}$ depends explicitly on $q$. We note that for large values of $q$, the power-law exponent $\beta_{\rm IKZM}(q)$ reduces to that in homogeneous systems.

The power law in Eq. (\ref{qdikzm}) is the analogue in spatially inhomogeneous systems of the expression for  the density of defects generated in the passage through a quantum critical point induced by a nonlinear quench in the time domain \cite{Diptiman,Barankov}. Specifically, the latter scenario concerns a homogeneous system driven by a homogeneous field such that $\epsilon= |t/\tau_Q|^r$, and leads to density of defects $d\propto\tau_Q^{-\frac{r \nu}{1+r z\nu}}$.

We further note that inhomogeneous driving results in a power-law suppression of the density of defects with respect to the homogeneous scenario
\beqa
\frac{d_{\rm IKZM}}{d_{\rm KZM}}=\frac{2}{L}\left(\frac{\hbar}{2\alpha_q q\tau_QJ(0)}\right)^{\frac{1}{q-1}}.
\eeqa

\subsection{A. Density of defects with arbitrary spatial dependence of the critical front and critical exponents $\nu$ and $z$}

The preceding section can be generalized for an arbitrary critical system characterized by a correlation length critical exponent $\nu$ and a dynamic critical exponent $z$.
As discussed in the main text, the expression for the relevant sound velocity, is then
\beqa 
\label{hats}
\hat s =  \frac {\hat \xi} {\hat \tau} = \frac {\xi_0} {\tau_0} \bigg[\frac {\tau_0} {\tau_Q(n)} \bigg]^{\frac {\nu(z-1)} {1+\nu z}}.
\eeqa  
Assuming defect formation to be restricted to the neighborhood of $n=0$, one can set  $\tau_{Q}(n)\approx\tau_Q(0)=\tau_Q$ in this expression.
The condition $v_F>\hat s $ leads to the estimate of (half) the effective size for defect formation
\beqa
|n|<|\hat{n}|=\left(\frac{1}{\alpha_q q \xi_0}\right)^{\frac{1}{q-1}}\bigg(\frac {\tau_0} {\tau_Q} \bigg)^{\frac {\nu(z-1)} {(1+\nu z)(q-1)}}.
\eeqa
Using the KZM estimate for the size of the domains, $\hat{\xi}=\xi_0(\tau_Q/\tau_0)^{\nu/(1+\nu z)}$, the density of defects becomes
\beqa
d_{\rm IKZM}=\frac{2|\hat{n}|}{L\hat{\xi}}=\frac{2}{L\xi_0}\left(\frac{1}{\alpha_q q \xi_0}\right)^{\frac{1}{q-1}}\bigg(\frac {\tau_0} {\tau_Q} \bigg)^{\frac {1+q\nu} {(1+\nu z)(q-1)}}.
\eeqa

\subsection{B. Deviations from power-law behavior: quantum Ising model}
The case with $q<1$ is not without interest as 
\beqa
v_F=\alpha_q q\tau_Q|n|^{1-q},
\eeqa

this is, it increases with $\tau_Q$ and away from the center of the chain $n=0$.
The condition for topological defect formation is then fulfilled in two disconnected regions, $[-L/2,-\hat{n}]$ and $[\hat{n},L/2]$,
where
\beqa
|\hat{n}|>\left(\frac{2J(n)}{\hbar\alpha_q q\tau_Q}\right)^{\frac{1}{1-q}}.
\eeqa

Thus, the spatial distribution of topological defects is expected to  be concentrated at the edges of the chain as opposed to its center.
In turn, the predicted density of topological defects  no longer follows a simple power-law scaling
\beqa
d_{\rm IKZM}&=&\frac{2(L/2-|\hat{n}|)}{L\hat{\xi}}\\
&=&d_{\rm KZM}-\frac{2}{L}\left(\frac{4J(n)J(0)}{\hbar^2\alpha_q q}\right)^{\frac{1}{1-q}}\left(\frac{\hbar}{2J(0)\tau_Q}\right)^{\frac{3-q}{2-2q}}.\nonumber
\eeqa
being governed by the combination of two different power-laws.

The lack of a power-law scaling can also occur for $q>1$ whenever defect formation is not restricted to the neighborhood of $n=0$. To appreciate this, it is required to take into account the spatial modulation of $s(n)$ and note that the condition  $v_F>s(n)$ can then be satisfied both in the proximity of $n=0$ as well as near $n=\pm L/2$.
The condition for defect formation, $v_F>s(n)$, then yields
\beqa
\label{effsize}
|n|^{q-1}(1-\alpha_{q}|n|^q)<\frac{\hbar}{2\alpha q\tau_QJ(0)}
\eeqa
The preceding analysis  follows from disregarding the second term in the left hand side, that leads to deviations from \eqref{qdikzm}.
\section{II. Numerical Approach}
\subsection{A. Numerical Methods: DMRG and TEBD} 
We perform numerical simulations based on tensor-network algorithms. The numerical results were obtained using the Tensor Network Theory (TNT) open-source library~\cite{sarah}. This library contains highly optimized routines for several algorithms based on a tensor network structure. \\
\\
{\it Initialization in Simulations of  Inhomogeneous Quantum Dynamics.---} We considered a time-dependent Hamiltonian describing a spin chain with inhomogeneous short-range spin-spin interactions $J\pap{n}$, where the tunable longitudinal matrix element $J\pap{n}$ is  modulated smoothly as a function of the site index $n$, i.e.,  as $J\pap{n}=J_{0}\pap{1-\alpha_{q}|n|^{q}}$ with $\alpha_{q}>0$ and $q>1$. The system Hamiltonian is       
\begin{equation}\label{Hannel}
\hat{\mathcal{H}}\pap{t}=-\sum_{n=1}^{L-1}J\pap{n}\hat{\sigma}_{n}^{z}\hat{\sigma}_{n+1}^{z} - \sum_{n=1}^{L}h(t)\hat{\sigma}_{n}^{x},
\end{equation}
where $\hat{\sigma}_{n}^{\alpha}$ are the Pauli matrices acting on the $n$-th spin, along the $\alpha=\pac{x,z}$ direction. Additionally, the  magnetic field is taken to be homogeneous and varying with constant rate, 
\beqa
h(t)=J(0)\left(1-\frac{t}{\tau_Q}\right).
\label{rampeq}
\eeqa 

The  quantum dynamics starts at $t=-\tau_Q$.  We first calculate the ground state for~\eqref{Hannel} by means of the density matrix renormalization group (DMRG) algorithm~\cite{white,white1}, using a matrix-product state (MPS) and matrix-product  operator (MPO) description of the system with open boundary conditions~\cite{Scholl}. In  Fig.~\ref{Sfig_2}, we depict a schematic representation of DMRG algorithm. The input for the DMRG algorithm is a random MPS represented graphically by $\ket{R}$ and the MPO corresponding to Eq.~\eqref{Hannel} at $t=-\tau_Q$. The algorithm uses a traditional DMRG minimization process and yields the MPS for the  ground-state  and  the corresponding ground-state energy. 

\begin{figure*}[h!]
\begin{center}
\includegraphics[scale=1]{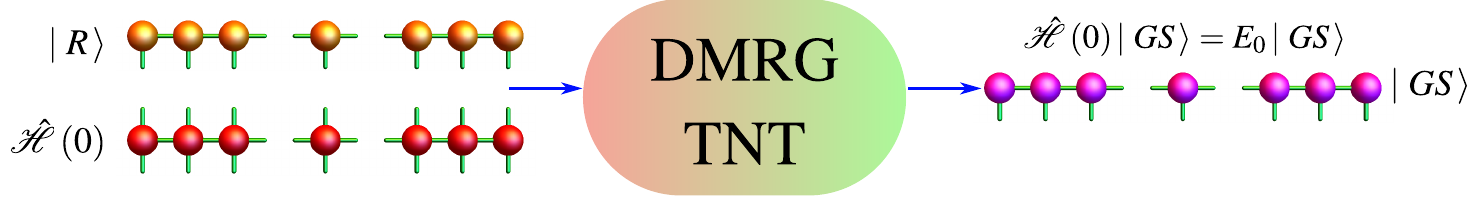}
\caption{{\bf Density Matrix Renormalization Group.} Schematic illustration of numerical DMRG protocol used to calculate the instantaneous ground state energy $\left|GS\right\rangle$ and  the corresponding ground-state energy $E_0$. The numerical routine uses as an input a random matrix-product state $\left| R\right\rangle$ and the matrix-product operator corresponding to Eq.~\eqref{Hannel} at $t=-\tau_Q$.}
\label{Sfig_2}
\end{center}
\end{figure*}
\begin{figure*}[h!]
\begin{center}
\includegraphics[scale=0.7]{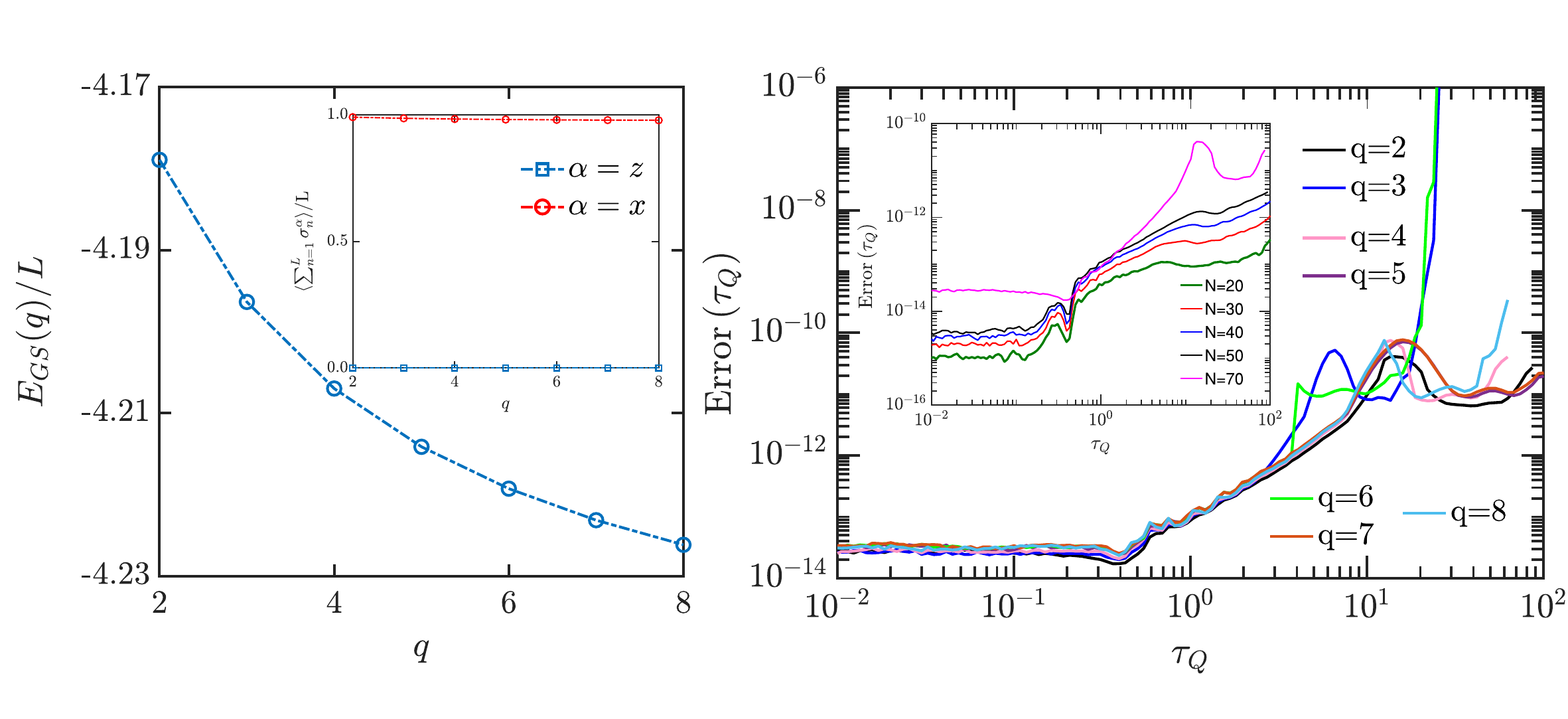}
\caption{{\bf Numerical test.} Left: The main panel show the expectation value of ground state energy as a function of $q$. In the inset, the expectation value of magnetization is shown along $x$ and $z$ direction. The system size is $L=70$. Right: In the main panel, the final truncation error  is shown as a function of $\tau_Q$, for severals values of $q$. The inset  shows the final truncation error for different systems sizes for $q=2$.}
\label{Sfig_3}
\end{center}
\end{figure*}

\newpage
Figure~\ref{Sfig_3} shows the ground state energy per particle as a function of $q$, as well as the initial magnetization along $x$ and $z$ direction. For this DMRG process, we consider MPO dimension up to $\chi=500$.     
\\
The time evolution induced by  the ramp across the critical point (\ref{rampeq}) is implemented by the 
 Time Evolving Decimation Block (TEBD) algorithm, suited  for one-dimensional many-body systems. 
 During the simulation of non-equilibrium dynamics,  correlations in the system generally increase with the time of evolution and  
convergence is checked by increasing the MPO dimension $\chi$. In the main right panel of Fig.~\ref{Sfig_3}, we show the final truncation error,  calculated as the sum of the truncation errors of each singular value decomposition performed.  

\subsection{B. Numerical results  with arbitrary spatial dependence of the critical front: quantum Ising model}\label{Ape3}

Figure \ref{Sfig_2} shows the scaling of the density of defects as a function of the quench time $\tau_Q$ for different values of $q$ governing the spatial dependence of the tunneling matrix elements $J(n)$. In all cases, the density of defects saturates to a constant value in the limit of the fast quenches. For slower values, a scaling law is first observed that is well described by the KZM for homogeneous systems. In this regime the velocity fo the front $v_F$ exceeds the relevant speed of sound everywhere in the system making the transition effectively homogeneous, in spite of the spatial dependence  of $J(n)$. For even slower quenches, a second scaling regime is observed where the power-law exponent is in agreement with the prediction in Eq. \eqref{qdikzm}. This is the main quantitative prediction of the KZM extension to inhomogeneous systems in both classical and quantum domains.  Interestingly, the crossover between the homogeneous and inhomogeneous scenario is clearer for even values of $q=4,6,8$ when the density of defects drops a fine value across the crossover. The onset of adiabatic dynamics is found for even slower values of the quench time. We emphasize that deviations from the scaling regimes illustrated here can be expected whenever defect formation not restricted to the center of the chain, in view of Eq. \eqref{effsize}. The fitted power-law exponents are collected in Table \ref{qtable} for both the homogeneous and inhomogenous power-laws.
While the exponent $\beta_{\rm KZM}$ approaches the constant theoretical value  $\beta_{\rm KZM}=1/2$, the larger exponent $\beta_{\rm IKZM}$ exhibits a dependence on the value of $q$ in agreement with the prediction $\beta_{\rm IKZM}=(q+1)/(2q-2)$.
\begin{figure}
\begin{tabular}{| c | c | c || c | c || c |}\hline\hline
$q$ & $\beta_{\rm KZM}\pm\Delta\beta_{\rm KZM}$& $r^{2}_{\rm KZM}$ &  $\beta_{\rm IKZM}\pm\Delta\beta_{\rm IKZM}$& $r^{2}_{\rm IKZM}$ & $\beta_{\rm IKZM}(q)$\\ \hline\hline
$2$ & $0.52\pm 0.03$ & $0.9991$ & $1.51\pm 0.03$ & $0.9993$ &3/2\\\hline 
$3$ & $0.56\pm 0.03$ & $0.9990$ & $1.08\pm 0.10$ & $0.9993$ &1\\\hline
$4$ & $0.58\pm 0.02$ & $0.9991$ & $0.77\pm 0.07$ & $0.9993$ &5/6\\\hline 
$5$ & $0.57\pm 0.02$ & $0.9990$ & $0.70\pm 0.03$ & $0.9999$&3/4\\\hline
$6$ & $0.55\pm 0.02$ & $0.9990$ & $0.65\pm 0.05$ & $0.9991$&7/10\\\hline
$7$ & $0.62\pm 0.02$ & $0.9996$ & $0.64\pm 0.02$ & $0.9997$&2/3\\\hline
$8$ & $0.57\pm 0.03$ & $0.9992$ & $0.61\pm 0.01$ & $0.9996$&9/14\\\hline\hline
\end{tabular}
\caption{{\bf Numerical power-law exponents.} The density of defects generated across an inhomogeneous phase transition exhibits a crossover from a power-law  $d_{\rm KZM}\sim\tau_Q^{-\beta_{\rm KZM}}$ characteristic of truly homogeneous systems to a second power-law of the form  $d_{\rm IKZM}\sim\tau_Q^{-\beta_{\rm IKZM}}$, describing slower quenches with larger values of the quench time $\tau_Q$. We show the numerical results for the fitted corresponding power-law exponents $\beta_{\rm KZM}$ and $\beta_{\rm IKZM}$ for severals values of $q$ and system size $N=70$. The power-law exponent $\beta_{\rm KZM}$ is approximately independent of the value of $q$ and takes slightly higher values than the theoretical prediction $\beta_{\rm KZM}=1/2$ for the Ising model, in agreement with previous theoretical and numerical studies. The second power law is characterized by an exponent in good agreement with the theoretical prediction $\beta_{\rm IKZM}=(q+1)/(2q-2)$.}\label{qtable}
\end{figure} 

\begin{figure*}[t]
\begin{center}
\includegraphics[scale=0.87]{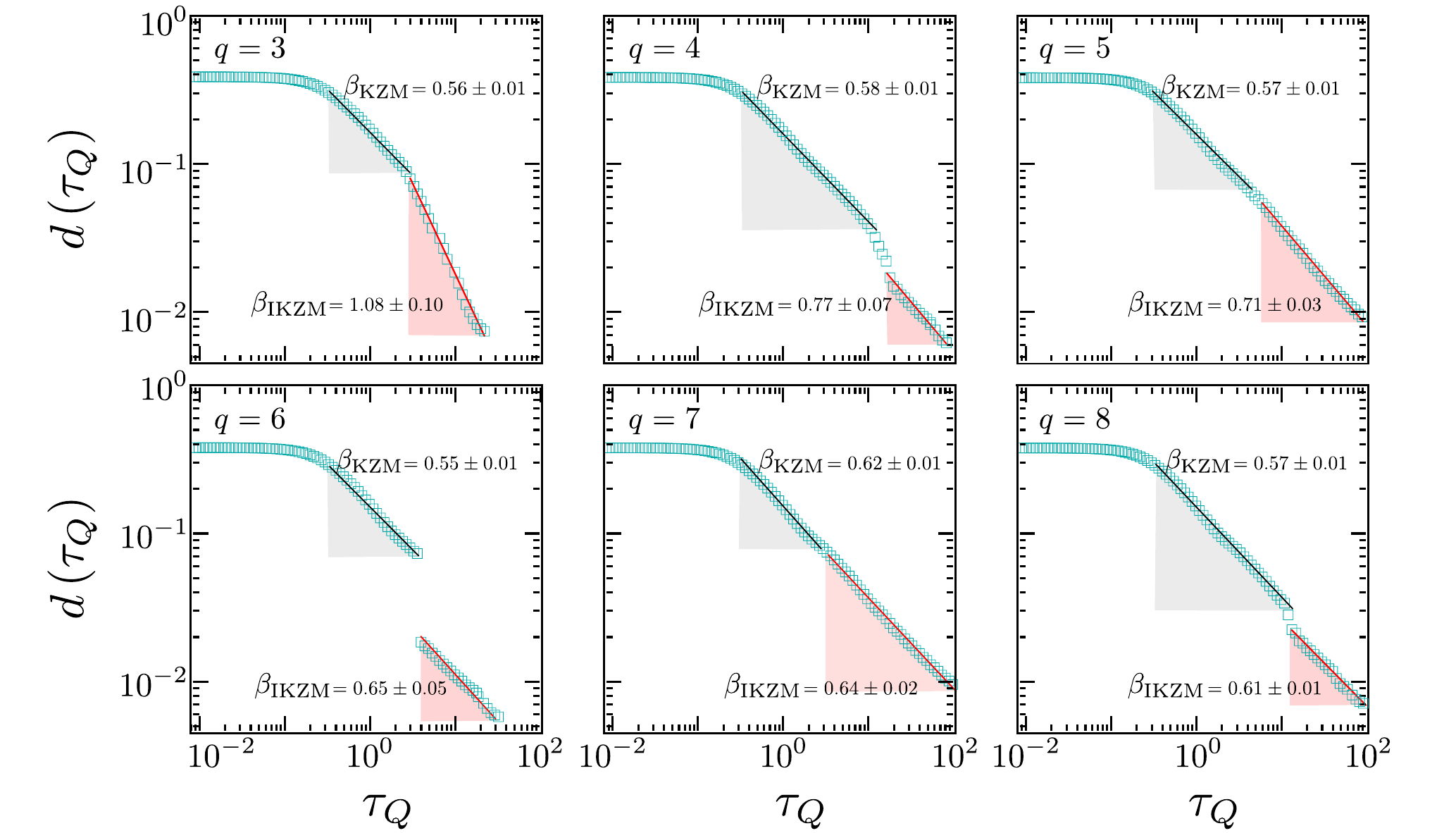}
\caption{{\bf Density of topological defect with  a non-quadratic tunneling matrix element $J(n)$.} The density of  defects is plotted as a function of the quench time in a doubly logarithmic scale for different values of $q$ and system size $N=70$. In all cases a transition is observed from a regime governed by homogeneous KZM at fast quenches and an enhanced suppression of defects for slow quenches. The latter arises from the interplay of the velocity of the critical front $v_F$ and the speed of sound $s$ and is the key prediction of the IKZM. }\label{Sfig_2}
\end{center}
\end{figure*}

\end{document}